\documentclass{PoS}

\usepackage{xspace}
\usepackage{amsmath} %Never write a paper without using amsmath for its many new commands 
\usepackage{amssymb} %Some extra symbols 

\newcommand{\nubar}{$\overline{\nu}\ $}
\newcommand{\nue}{\ensuremath{\nu_{e}}\xspace}
\newcommand{\nubare}{\ensuremath{\overline{\nu}_{e}}\xspace}

\newcommand{\numu}{\ensuremath{\nu_{\mu}}\xspace}
\newcommand{\nust}{\ensuremath{\nu_s}\xspace}

\newcommand{\nutau}{\ensuremath{\nu_{\tau}}\xspace}
\newcommand{\numunue}{\ensuremath{\nu_\mu \rightarrow \nu_e}\xspace}
\newcommand{\numunumu}{\ensuremath{\nu_\mu \rightarrow \nu_\mu}\xspace}
\newcommand{\numunust}{\ensuremath{\nu_\mu \rightarrow \nu_s}\xspace}
\newcommand{\nuenue}{\ensuremath{\nue \rightarrow \nue}\xspace}

\newcommand{\numunutau}{\ensuremath{\numu \rightarrow \nutau}\xspace}
\newcommand{\nuenutau}{\ensuremath{\nue \rightarrow \nutau}}

\newcommand{\boss}[2]{\ensuremath{\rlap{\kern-2.5pt\ensuremath{\overset{\scriptscriptstyle(-)}{\phantom{#1}}}}{\ensuremath{{#1}_{#2}}}}}

\title{Search for Sterile Neutrinos at OPERA and other Long--Baseline Experiments}

\ShortTitle{Sterile Neutrino at LBL}

\author{\speaker{Luca Stanco}\thanks{On behalf of the OPERA Collaboration.}\\
        I.N.F.N. Padova (Italy)\\
        E-mail: \email{stanco@pd.infn.it}}

\abstract{The OPERA experiment at the CNGS beam has observed muon to tau neutrino oscillations in the atmospheric sector. 
Based on this result new limits on the mixing parameters of a massive sterile neutrino may be set. 
Preliminary results of the analysis done in the $3+1$ neutrino framework are here presented.\\ 
An update of the search for sterile neutrinos in the \numunue channel is also given. \\
The status of the searches for sterile neutrinos performed at other Long--Baseline experiments (MINOS at NuMI beam and SuperK with the atmospheric flux) is also briefly recalled. Finally,  some personal perspectives are discussed.}

\FullConference{The European Physical Society Conference on High Energy Physics\\
		22--29 July 2015\\
		Vienna, Austria}

\begin{document}

\section{Introduction}\label{sect-1}

The present scenario of the Standard Model (SM) for particle physics, being, so to say, suspended by the discovery of the Higgs boson,
is desperately looking for new experimental inputs to provide a more comfortable and conformable theory.
From another point of view, experiments on neutrinos have been so far an outstanding source of novelty and unprecedented 
results. In the latest two decades several results were obtained from the study of neutrinos, either from
the atmospheric, solar and reactor ones or, more recently, from accelerator--based beams. Almost all these results contributed to strengthen  a wonderful fullfilment of the flavour--SM.

On the other hand, some of the neutrino results (fortunately) concern anomalies that do not fit in the standard scenario. In particular three
different kinds of experiments hint at the possible existence of at least a 4$^{\rm{th}}$ and \emph{sterile} (i.e. not coupled to the weak interaction) neutrino. The excess
of \nue (\nubare) observed by the LSND~\cite{lsnd} and MiniBooNE~\cite{miniB} collaborations and the so-called 
reactor~\cite{reactor} and Gallium~\cite{gallium1, gallium2} neutrino anomalies can be coherently interpreted as due to the existence of a fourth sterile neutrino with a mass at the eV scale.

There are presently many proposals and experiments~\cite{next-prop} that tentatively address the issue, and in the next few years results
are expected to confirm/disprove the previous anomalies. However, to the conviction of the author, none of them will
be able to establish the existence/interpretation of the results in terms of sterile neutrinos. In fact, any extension of the 
3--flavour model, which is on the contrary perfectly adequate to the ``standard'' results, owns internal strong tensions between 
the interpretation of \nue (\nubare) appearance/disappearance and the corresponding, required, appearance/disappearance
of the other flavours, \numu and \nutau ~\cite{tension}.

Therefore it would be imperative to search for anomalies related to the appearance/disappearan\-ce of \numu and \nutau
neutrinos. Following this approach there are investigations pursued with neutrino accelerator beams,
even if only (unfortunately) at Long--Baseline (LBL). In this paper we will present the very recent, preliminary result 
on the \nutau non--standard appearance/disappearance, following the observation of the 5$^{\rm{th}}$ \nutau candidate 
by the OPERA experiment~\cite{opera-5th}. After a description of the oscillation constraints at LBL (section~\ref{sect-2}), we will focus 
on the OPERA analyses, also including some very recent results in the \numunue channel (section~\ref{sect-3}). In section~\ref{sect-4} reports on the MINOS and SuperK analyses will be shortly described.
Conclusions and perspectives will be finally drawn (section~\ref{sect-5}).

\section{The Long-Baseline scenario for sterile neutrinos}\label{sect-2}

The presence of an additional sterile--state can be expressed in the extended PMNS~\cite{pmns} mixing matrix 
($U_{\alpha i}$ with $\alpha = e, \mu, \tau, s$, and $i = 1,\ldots,4$). 
In this model, called ``3+1'', the neutrino mass eigenstates $\nu_1,\ldots,\nu_4$ are labeled such that the first three states are mostly made 
of active flavour states and contribute to the ``standard'' three flavour oscillations with the squared mass differences 
$\Delta\, m_{21}^2 \sim 7.5\times 10^{-5}~{\rm eV^2}$ and $|\Delta\, m_{31}^2| \sim 2.4\times 10^{-3}~{\rm eV^2}$, 
where $\Delta\, m_{ij}^2 = m^2_i - m^2_j$. The fourth mass eigenstate, which is mostly sterile, is assumed to be much heavier than the others, 
$0.1~{\rm eV^2}\lesssim \Delta\, m_{41}^2 \lesssim 10~{\rm eV^2}.$ The opposite case in hierarchy, i.e. negative values of $\Delta\, m_{41}^2$, produces a similar phenomenology from the 
oscillation point of view but is disfavored by cosmological results on the sum of neutrino masses~\cite{cosmo-data}. 

In a Short-Baseline experiment the oscillation effects due to $\Delta\, m^2_{21}$ and $\Delta\, m^2_{31}$ can be neglected since $L/E\sim 1$ km/GeV. 
Therefore the oscillation probability depends only on $\Delta\, m^2_{41}$ and $U_{\alpha 4}$ with $\alpha = e,\mu,\tau$.
In particular the survival probability of muon neutrinos can be given by an effective two--flavour oscillation formula.

Differently when $L/E \gg 1$ km/GeV, that is the case for the Long-Baseline experiments, the two--flavour oscillation
is not a good approximation. In the case of the CNGS beam, when studying the
\nutau oscillation rate the only valid approximations correspond to neglect
 the solar--driven term, i.e. $\Delta\, m_{21}^2 \sim 0$,  and to discard  the \nue component of
the beam. However when the \numunue channel is studied the intrinsic \nue beam--component becomes 
a non--negligible factor~\cite{palazzo}.

Considering the (\numu, \nutau, $\nu_s$) triplet, together with the above two approximations, the most general oscillation probability \numunutau can be written as:
\begin{eqnarray*}{P_{\nu_\mu\to\nu_{\tau}} = {4 |U_{\mu 3}|^2|U_{\tau 3}|^2\sin^2\frac{\Delta_{31}}{2}} + {4 |U_{\mu 4}|^2|U_{\tau 4}|^2\sin^2\frac{\Delta_{41}}{2}}}\\
{ +\, { 2\Re[U^*_{\mu 4}U_{\tau 4}U_{\mu 3}U_{\tau 3}^*]\sin\Delta_{31}\sin\Delta_{41}}}\\
{ -\, { 4\Im[U^*_{\mu 4}U_{\tau 4}U_{\mu 3}U_{\tau 3}^*]\sin^2\frac{\Delta_{31}}{2}\sin\Delta_{41}}}\\
{ +\, { 8\Re[U^*_{\mu 4}U_{\tau 4}U_{\mu 3}U_{\tau 3}^*]\sin^2\frac{\Delta_{31}}{2}\sin^2\frac{\Delta_{41}}{2}}}\\
{+\, {4 \Im[U^*_{\mu 4}U_{\tau 4}U_{\mu 3}U_{\tau 3}^*]\sin\Delta_{31}\sin^2\frac{\Delta_{41}}{2}}},
\end{eqnarray*}
using the definition $\Delta_{ij}=1.27\; \Delta\, m^2_{ij}\; L/E$ ({\em i,j=1,2,3,4}), with $\Delta_{31}$ and $\Delta_{41}$
expressed in eV$^2$, $L$ in km and $E$ in GeV. The first term corresponds to the standard oscillation, the
second one to the pure exotic oscillation, while the next 4 terms correspond to the interference between the standard
and sterile neutrinos. By defining $C=2|U_{\mu3}||U_{\tau3}|$, 
$\phi_{\mu\tau}=Arg(U^{\star}_{\mu3}U^{\star}_{\tau3}U^{\star}_{\mu4}U^{\star}_{\tau4})$ and 
$\sin\, 2\theta_{\mu\tau}=2|U_{\mu4}||U_{\tau4}|$ the expression can be re--written as:
\begin{eqnarray*}
{P(Energy) = {\color{black} C^2 \sin^2\frac{\Delta_{31}}{2}} + {\sin^2\, 2\theta_{\mu\tau}\sin^2\frac{\Delta_{41}}{2}}}\\
{ +\, { \frac{1}{2} C \sin\, 2\theta_{\mu\tau}\cos\phi_{\mu\tau}\sin\Delta_{31}\sin\Delta_{41}}}\\
{ -\, { C\sin2\theta_{\mu\tau}\sin\phi_{\mu\tau}\sin^2\frac{\Delta_{31}}{2}\sin\Delta_{41}}}\\
{ +\, { 2\, C\sin2\theta_{\mu\tau}\cos\phi_{\mu\tau}\sin^2\frac{\Delta_{31}}{2}\sin^2\frac{\Delta_{41}}{2}}}\\
{+\, {C\sin 2\theta_{\mu\tau}\sin\phi_{\mu\tau}\sin\Delta_{31}\sin^2\frac{\Delta_{41}}{2}}},
\end{eqnarray*}
where interesting dependences rise up, namely the sign of $\Delta\, m^2_{13}$ (3$^{rd}$ and 6$^{th}$ terms) and 
the $\phi_{\mu\tau}$ CP-violating phase (4$^{th}$ and 6$^{th}$ terms). Finally, since at LBL $\sin\Delta_{41}\approx 0$
and $\sin^2\frac{\Delta_{41}}{2} \approx \frac{1}{2}$, the following expression is obtained:
\begin{eqnarray*}
{P(Energy) \simeq { C^2 \sin^2\frac{\Delta_{31}}{2}} + {\frac{1}{2}\sin^22\theta_{\mu\tau}}}\\
{ +\, { C\sin2\theta_{\mu\tau}\cos\phi_{\mu\tau}\sin^2\frac{\Delta_{31}}{2}}}\\
{+\, {\frac{1}{2}C\sin\, 2\theta_{\mu\tau}\sin\phi_{\mu\tau}\sin\Delta_{31}}}.
\end{eqnarray*}
This formula indicates that we are sensitive to the effective sterile mixing angle, $\theta_{\mu\tau}$, the mass hierarchy (MH, Normal NH or Inverted IH) and to the CP--violating phase. 

The method carried out by the OPERA collaboration is independently applied to 
the NH and IH cases. Maximization of the likelihood is performed
over the CP-violation phase, $\phi_{\mu\tau}$, and the two effective mixing angles of the 3$^{rd}$ and $4^{th}$ 
mass--states 
with \numu and \nutau, the variables $C$ and $\theta_{\mu\tau}$, respectively. 

\section{OPERA preliminary results on sterile neutrinos from \nutau and \nue}\label{sect-3}

\begin{figure}
     \includegraphics[width=.5\textwidth]{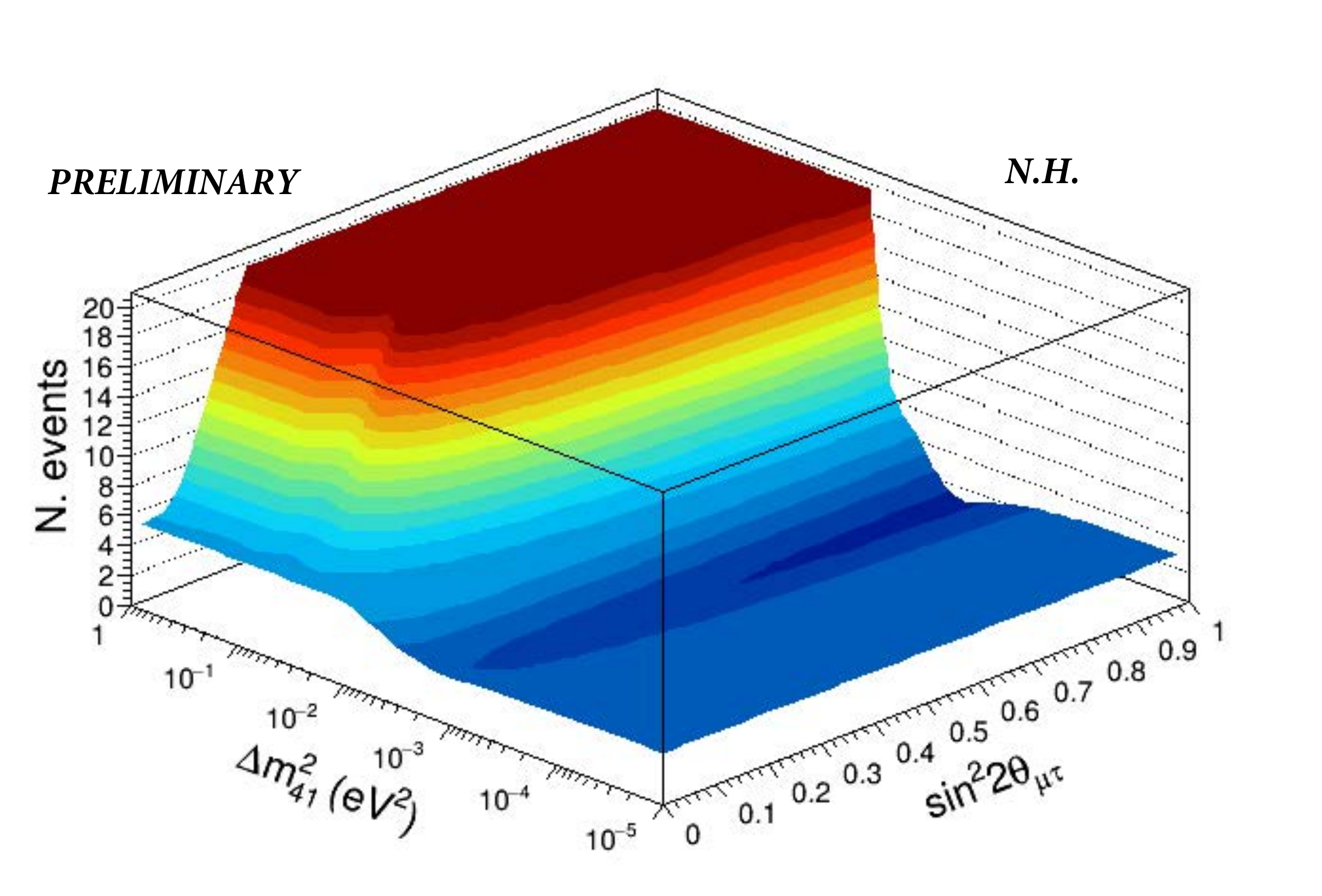}
     \includegraphics[width=.5\textwidth]{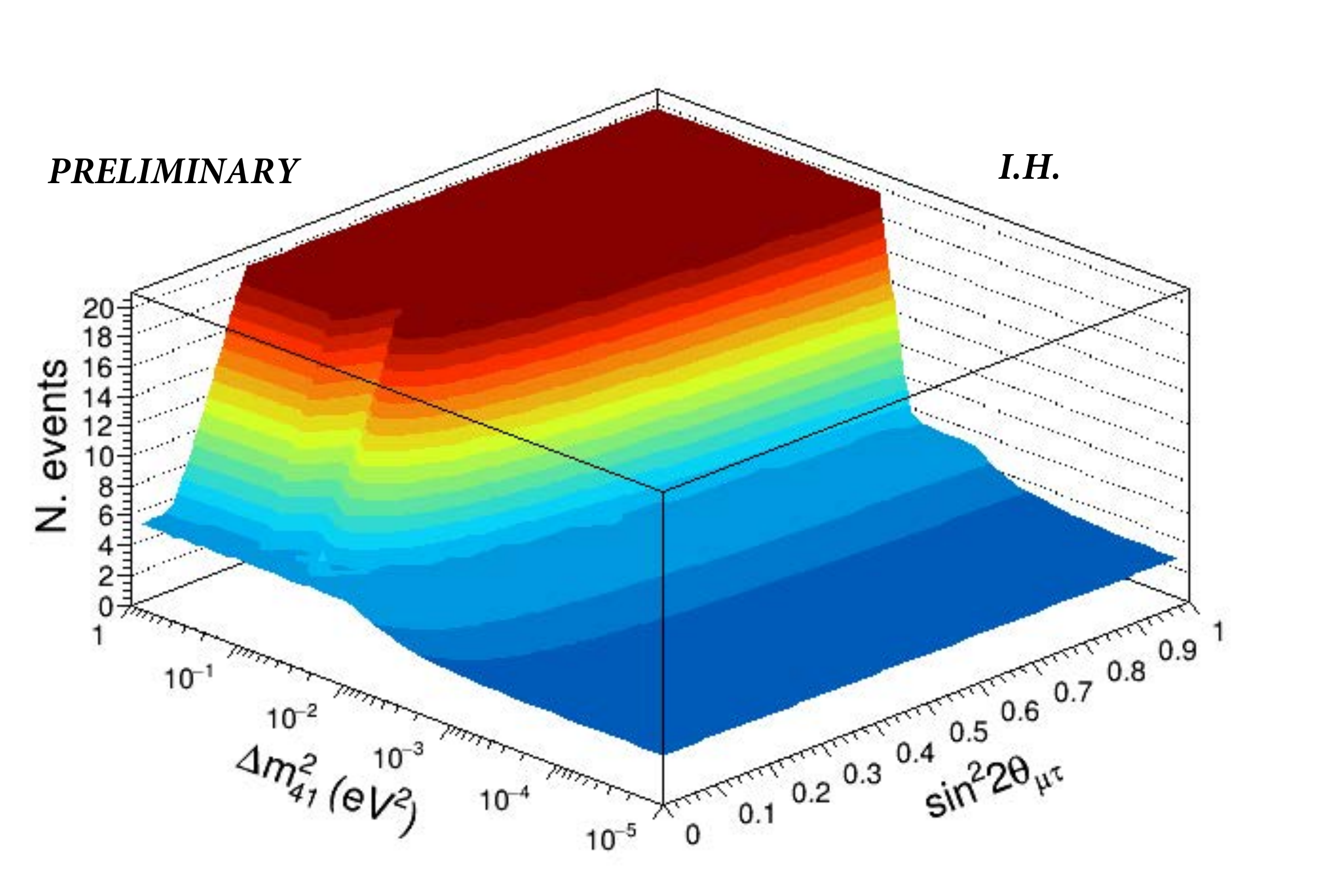}
     \caption{The expected increase/decrease of events for the 3+1 model, as function of $\Delta\, m^2_{41}$ and the effective
mixing angle, $\theta_{\mu\tau}$, of \numu and \nutau with the 4$^{\rm{th}}$ neutrino state, and maximization
over the other parameters.}
     \label{fig1}
\end{figure}

Results on sterile limits based on four \nutau candidates have been very recently published by OPERA~\cite{opera-4th}. We report here the preliminary
updated analysis based on the just discovered 5$^{th}$ candidate~\cite{opera-5th}.
In figure~\ref{fig1} the net result on the expected number of \nutau candidates 
is shown in presence of a 4$^{\rm{th}}$ neutrino state, as a function of $\sin^2\, 2\theta_{\mu\tau}$ and $\Delta\, m^2_{41}$.

\begin{figure}\begin{center}
     \includegraphics[width=.6\textwidth]{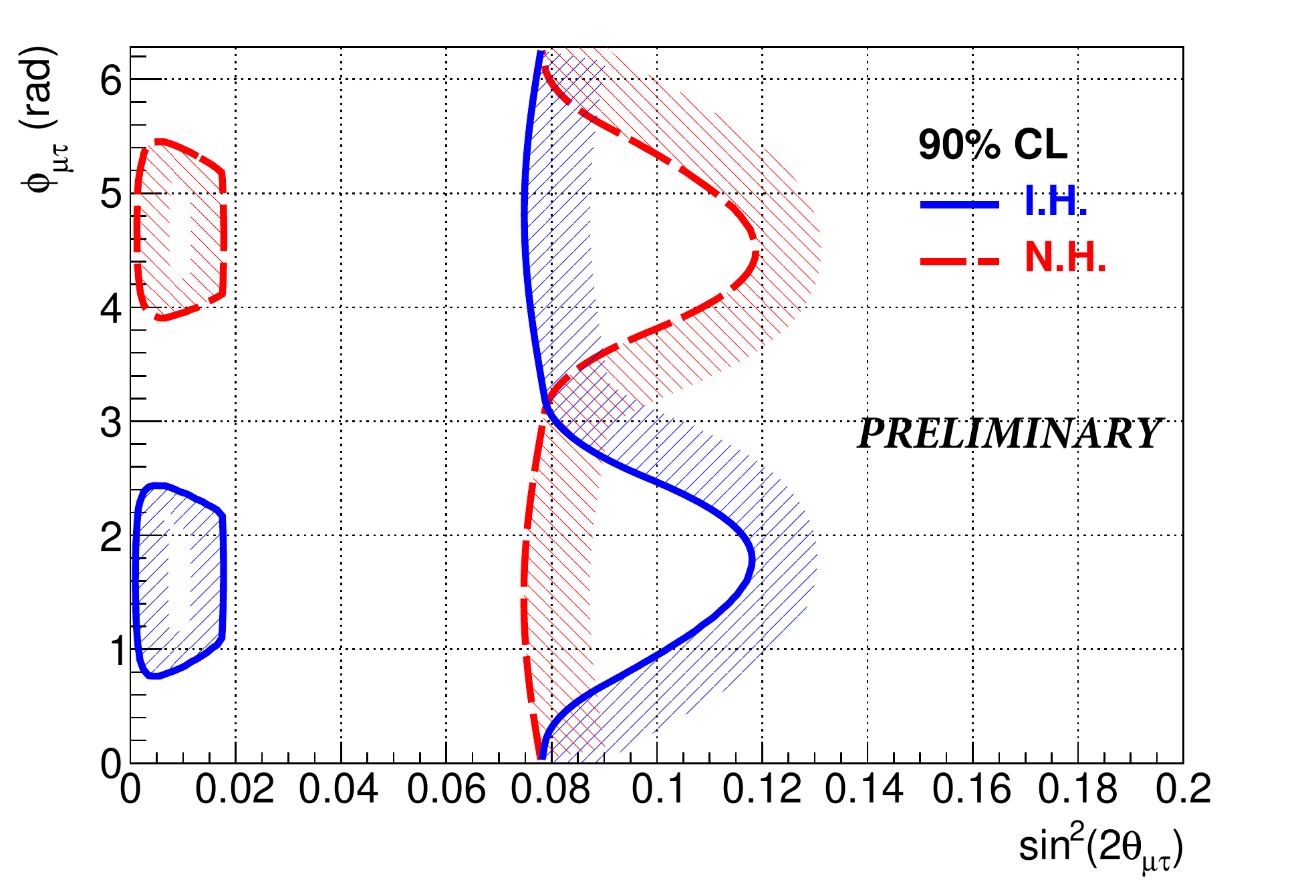}
     \caption{90\% C.L. exclusion limits in the $\phi_{\mu\tau}$ vs $\sin^2\, 2\theta_{\mu\tau}$ parameter space for normal 
(NH, dashed red) and inverted (IH, solid blue) hierarchies, and assuming $\Delta\, m_{41}>1$ eV$^2$. Bands are drawn to 
indicate the excluded regions.}
     \label{fig2}
\end{center}\end{figure}

From the picture it is evident that OPERA sensitivity is limited to the region ($\sin^2\, 2\theta_{\mu\tau}\gtrsim 0.1$, 
$\Delta\, m_{41}\gtrsim 0.01$ eV$^2$). Therefore the OPERA analysis was two--fold. In the first case the
$\Delta\, m_{41}~>~1$~eV$^2$ region was considered, where almost no correlation with the effective mixing angle is 
present. Then the exclusion limit on the plane of the phase vs the mixing angle  can be extracted (figure~\ref{fig2}).
When marginalization over the phase is made, the limit $\sin^2\, 2\theta_{\mu\tau}< 0.11$ at 90\% C.L. is obtained
(preliminary).

\begin{figure}\begin{center}\vskip-0.7cm
     \includegraphics[width=.7\textwidth]{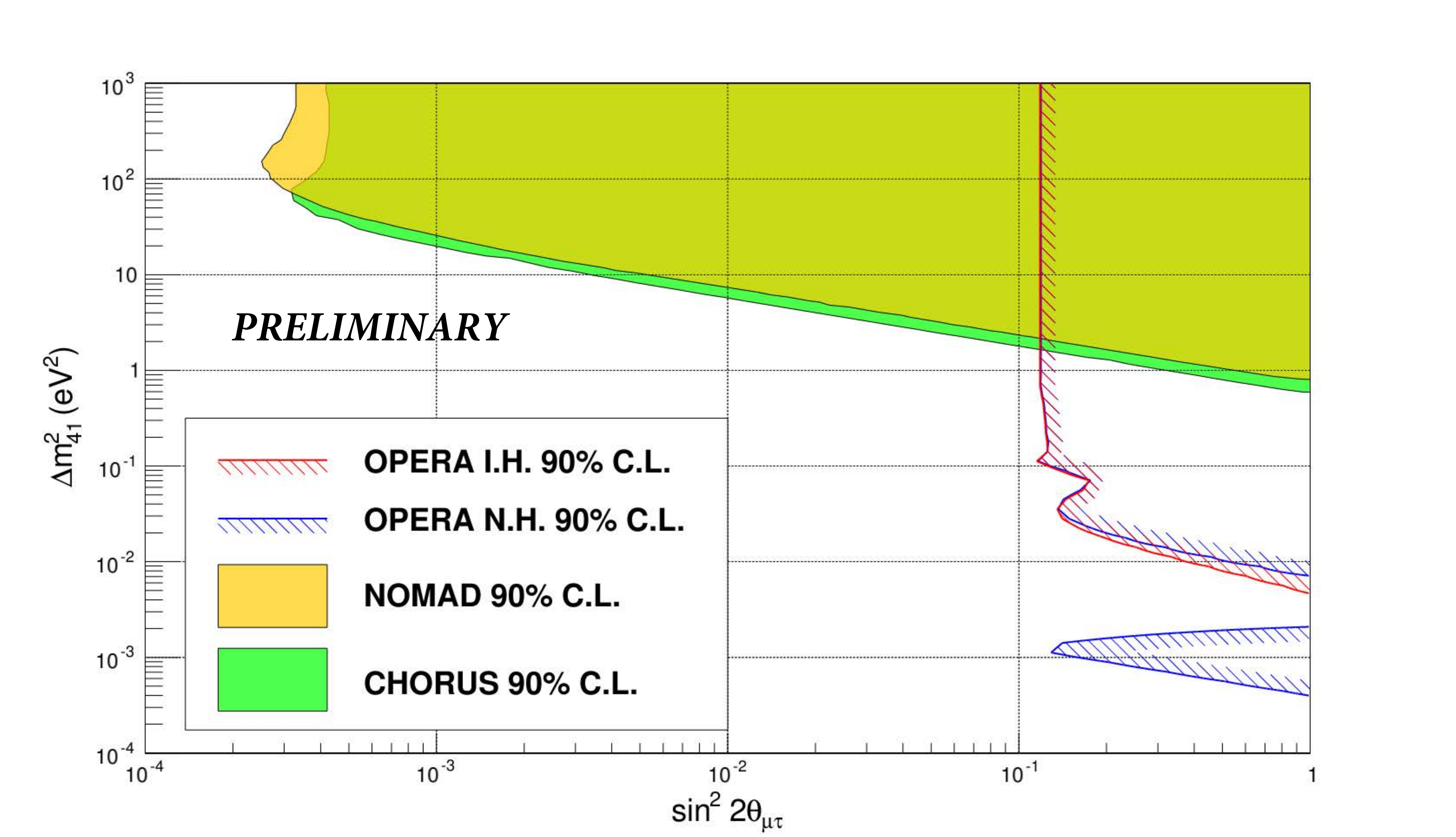}
     \caption{OPERA preliminary 90\% C.L. exclusion limits in the $\Delta\, m^2_{41}$ vs $\sin^2\, 2\theta_{\mu\tau}$ parameter space for
the normal (NH, red) and inverted (IH, blue) hierarchy of the three standard neutrino masses. The exclusion 
plots by NOMAD~\cite{nomad} and CHORUS~\cite{chorus} are also shown. Bands are drawn to indicate the excluded 
regions.}
     \label{fig3}
\end{center}\end{figure}

To extend the search for a possible fourth sterile neutrino down to small $\Delta\, m^2_{\mu\tau}$ values, a second
kind of analysis has been performed by OPERA using the GLoBES software,
which takes into account the non-zero $\Delta\, m^2_{12}$ value and also matter effects, the Earth density being approximated by a constant value estimated with the PREM shell--model.
This time the $\Delta\, m^2_{31}$ parameter has been profiled out (see~\cite{opera-4th} for more details and references).
In figure~\ref{fig3} the preliminary 90\% CL exclusion plot is reported in the $\Delta\, m^2_{41}$ vs $\sin^2\, 2\theta_{\mu\tau}$ 
parameter space. The most stringent limits of direct searches for \numunutau oscillations at short-baselines obtained 
by the NOMAD~\cite{nomad} and CHORUS~\cite{chorus} experiments are also shown.

Another very interesting analysis is ongoing, in progress in OPERA on the \numunue search. The 2013 analysis will be updated
using the full data set and a much less approximate analysis as previously done. The preliminary selection is shown in 
table~\ref{tab1}. The exclusion region will be set in the plane $\Delta\, m^2_{41}$ vs $\sin^2\, 2\theta_{\mu e}$.
\begin{table}[h]
\begin{center}\vskip-0.5cm
     \begin{tabular}{|l|c|c|}\hline
      & all energy range & $E<20$ GeV \\ \hline
     \nue candidates (30\% data) & 19 & 4\\ \hline
     \nue candidates (all data) & 52 & 9\\ \hline
     \end{tabular}
\end{center}
     \caption{The preliminary selected \nue candidates by OPERA.}
     \label{tab1}
     \end{table}

\section{MINOS and SuperK analyses}\label{sect-4}

The MINOS and SuperK collaborations have also studied in detail the \numunumu and \nuenue oscillations to 
exclude extra contributions from \numunust oscillations. Recent results have been given by MINOS
that makes use of the NuMi beam at FNAL~\cite{minos}, and SuperK by using the atmospheric flux~\cite{superk}.
MINOS is also analyzing the \nubar running--mode data--sample and their updated analysis
on \numunue will be soon released. 

The SuperK analysis is two--fold, considering either  $|U_{e4}|=0$ with matter effects or the full PMNS
and discarding the matter effect. In the latter case a strong limit is obtained, $|U_{\mu4}|<0.04$ at 90\% C.L., for
$\Delta\, m^2_{41} > 0.1$ eV$^2$, for a total exposure of 282 kton--year.

\section{Conclusions and Perspectives}\label{sect-5}

The long--standing issue on the existence of sterile neutrino states at the eV mass scale can receive new relevant inputs 
from the accelerator Long--Baseline experiments, like OPERA, MINOS and SuperK. From one side LBL, owing
to the large $L/E$ values, can only look at the averaged oscillation pattern (lacking any oscillatory behavior of data).
From the other side, the not--negligible interference between flavours rises up dependencies on the mass hierarchy and the CP--violation phase. 

New results were recently
published by the three collaborations, either on \numunumu disappearance or on the \numunutau appearance. 
All the results put stringent exclusion limits on the effective mixing angles between \numu/\nutau and \nust, so increasing the
tension with the positive results on \nue appearance/disappearance. With regards to \nue, OPERA and MINOS$+$
will soon release reliable results with their large data--set, by properly taking into account the extended $3+1$ scenario.

In case of existence of a sterile neutrino at the eV mass--scale this situation points towards a rather low effective mixing angle,
of the order of 1\%, between sterile and the standard neutrino flavours. 
Therefore for any experiment/proposal aiming to provide new results it is mandatory to reach a sensitivity of that level.

The sterile neutrino story has so far been developed either by trying to establish the hints (each at 2--3 $\sigma$ level) on
\nue appearance/disappearance, or looking at flavour connected channels, like the \numu disappearance one.
Within the next 2--3 years experiments on reactors and with radioactive sources can confirm or disprove the
\nue anomalies, while there is presently no reliable experiment~\cite{nessie} looking at the interference at the level of 1\% 
mixing between sterile and {\em muon/tau}
neutrino states other than the LBL ones. These  however have no possibility to observe the oscillation pattern.
Therefore, new specific  experiments should be settled and approved, in case reactor/source current investigations would
provide positive results.

\end{document}